# High-fidelity tabletop nanoscopy enabled by non-linear spectral preconditioning

Yun Xie[1], Zhiyi Huang[2], Bianli Zhao[3], Youyang Zhou[4], Steve F. Shu[3]

1. College of Intelligent Robotics and Advanced Manufacturing, Fudan University, Shanghai, 200433, P.R. China

2. School of Electrical and Computer Engineering, The University of Sydney, Camperdown, NSW 2006, Australia

3. Department of Materials and Energy, Yunnan University, Kunming 650504, P.R. China

4. School of Physics and Astronomy, Yunnan University, Kunming 650500, P.R. China

## Abstract

Ptychography is a powerful lensless imaging technique that overcomes conventional numerical aperture limits to achieve diffraction-limited resolution. While routine at high-brilliance synchrotron facilities, its application to laboratory-scale sources is primarily limited by low photon flux. Under these conditions, the wide dynamic range of diffraction signals presents a critical bottleneck where detector bit-depth limitations hinder the simultaneous recording of low-frequency intensity and high-frequency details. Currently, most high-dynamic-range (HDR) imaging methods enforce strict radiometric linearity, assuming the fused intensity must be linearly proportional to the squared modulus of the wavefront to satisfy Poisson likelihood models. In this paper, we introduce a multi-scale non-linear fusion approach into the ptychographic pipeline, demonstrating that strict linearity is not a prerequisite for accurate reconstruction. This method mitigates the traditional trade-off between noise suppression and physical fidelity, enables robust imaging under strong dispersion, and significantly broadens the effective spectral bandwidth.

## 1 Introduction

Ptychography reconstructs complex wavefronts from overlapping diffraction patterns using iterative phase retrieval algorithms (Park et al. 2018; Wang et al. 2022; Nguyen et al. 2022; Park et al. 2023; Rodenburg and Faulkner 2004; Maiden and Rodenburg 2009; Thibault et al. 2008; Stefano Marchesini 2016). Unlike conventional microscopy, which is limited by the numerical aperture of physical lenses (Chapman and Nugent 2010; Ozcan and McLeod 2016; Pfeiffer 2018; Valzania et al. 2019), this computational approach effectively bypasses optical diffraction limits (Jagatap and Hegde 2019; Bangun et al. 2022; Huang et al. 2015; Huijts et al. 2020; Zheng et al. 2013; Tanksalvala et al. 2021; Maiden et al. 2011; Faulkner and Rodenburg 2004). Recent advances in laboratory X-ray sources, such as liquid-metal-jet and high-harmonic generation (HHG) extreme ultraviolet (XUV) systems, have increased brightness, facilitating their application in ptychography. However, the broad spectral bandwidth of these sources often requires monochromatic filtering to meet coherence requirements. This filtering reduces the available photon flux, compounding the challenge presented by the wide dynamic range of diffraction signals, where the intensity difference between the zero-order beam and high-frequency scattering can exceed six orders of magnitude. This condition leads to a critical acquisition trade-off, particularly in photon-starved regimes (Giewekemeyer et al. 2014; Takahashi et al. 2023; Clark et al. 2011; Giewekemeyer et al. 2011). Short exposures result in high-frequency details being obscured by detector noise (Martin et al. 2012; Seki et al. 2018; Godard et al. 2012), whereas prolonged exposures, which are required to improve the signal-to-noise ratio (SNR), cause saturation and the irreversible loss of low-frequency data (Liu et al. 2021; Jiang et al. 2018; Philipp et al. 2022). Consequently, high-dynamic-range (HDR) fusion via multi-exposure acquisition is an essential strategy to overcome detector bit-depth limitations and preserve the full spatial frequency spectrum (Kodgirwar et al. 2024; Liu et al. 2021; Suzuki et al. 2016; Dierolf et al. 2010).

Most current fusion strategies in ptychography rely on radiometric linearity. Standard approaches, such as linear integration or threshold-based stitching (Dierolf et al. 2010; Liu et al. 2021), ensure that the measured intensity remains proportional to the squared modulus of the wavefront. However, these methods propagate stochastic shot noise from low-flux measurements into the fused data, degrading the SNR of high-frequency fringes. To address this, probabilistic frameworks (Kodgirwar et al. 2024; Takahashi et al. 2023; Thibault and Guizar-Sicairos 2012; Wei et al. 2022) formulate fusion as a Maximum Likelihood Estimation (MLE) problem based on Poisson statistics. While theoretically sound, these parametric models are sensitive to experimental instabilities, such as beam fluctuations or detector artifacts. This sensitivity can cause model mismatch, which amplifies outliers instead of suppressing them. Alternatively, structural fusion techniques from computer vision, such as Structural Patch Decomposition (SPD) (Ma et al. 2017; Xu et al. 2022), offer another approach to noise suppression. Although effective for natural images, these methods impose a high computational burden ($\mathcal{O}(Np^2)$) and tend to over-smooth diffraction data. This smoothing effect removes weak scattering signals and distorts the global intensity envelope necessary for diffraction physics, making such methods largely unsuitable for phase retrieval.

This highlights a need for strategies that balance the efficiency of linear methods with the noise suppression capabilities of non-linear approaches. While perceptual fusion algorithms (Mertens et al. 2009) are standard in computational photography, their application to ptychography remains limited. A primary reason is the concern that non-linear modulation compromises the quantitative consistency required for algorithmic convergence (Kodgirwar et al. 2024; Thibault and Guizar-Sicairos 2012; Driel et al. 2015). Iterative phase retrieval typically assumes the measured intensity is linearly proportional to the squared modulus of the wavefront ($I \propto |\psi|^2$), satisfying Poisson likelihood models (Godard et al. 2012). Deviations from this linearity are thus conventionally treated as sources of error that introduce structural artifacts (Driel et al. 2015). We propose, however, that non-linear modulation can redistribute measurement reliability based on structural integrity. Rather than causing divergence, it can act as an implicit regularization that smooths the rugged loss landscape caused by stochastic noise. Strict adherence to statistical fidelity in current frameworks precludes this gradient modulation, ultimately restricting achievable resolution (detailed theoretical derivations are provided in Supplementary Note 1).

In this work, we introduce non-linear multi-exposure fusion from computational photography into coherent diffractive imaging. By integrating this multi-scale approach directly into the phase retrieval pipeline, we demonstrate that strict radiometric linearity is not an absolute prerequisite for ptychographic reconstruction. We establish that structural consistency supersedes strict statistical linearity under photon-starved conditions. In such regimes, the noise suppression capabilities of non-linear fusion effectively compensate for induced radiometric deviations. To support this framework, we provide a rigorous theoretical foundation within a unified optimization model. An analysis of the gradient descent dynamics reveals that the non-linear fusion weights operate as a spatially adaptive spectral preconditioner. This mechanism acts as an implicit regularizer that filters stochastic gradient noise rather than a mere source of signal distortion. We also delineate the operational boundaries within which non-linear modulation enhances reconstruction without inducing physical model mismatch. By evaluating spectral topology and background variance, we define a regularization regime that recovers high-frequency diffraction orders buried in noise while preserving the physical consistency necessary for robust convergence.

# 2 Results

## 2.1 Fusion operation

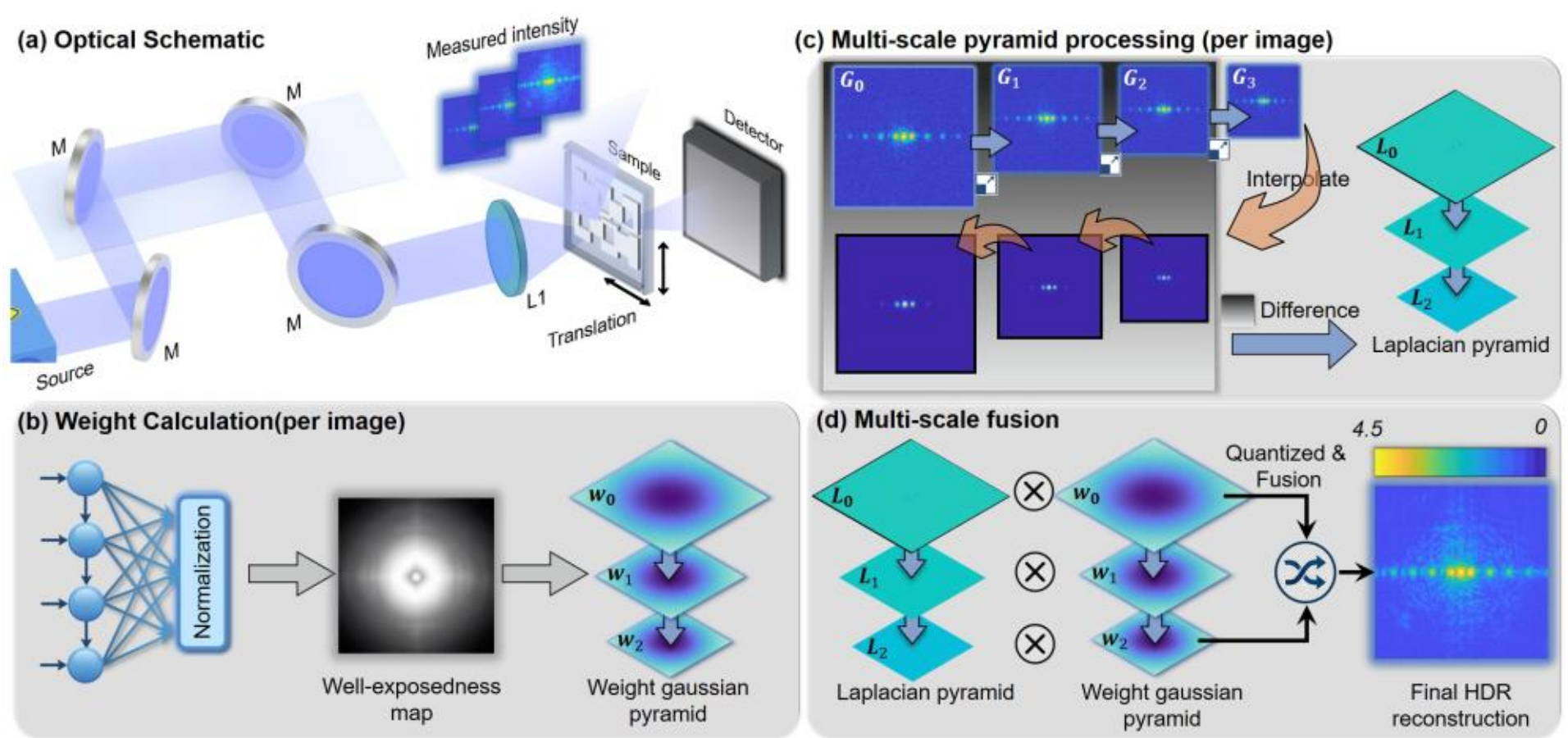

*Fig 1. Principle of Multi-Scale Non-Linear Fusion (MNF) operation. (a) Optical configuration of the ptychographic microscope. (b) A non-linear function assigns pixel-wise confidence scores based on local exposure quality. (c) Diffraction patterns and weight maps are transformed into Laplacian and Gaussian pyramids to isolate spatial frequency features. (d) The multi-scale features are modulated by local weights and collapsed to synthesize the final high-dynamic-range (HDR) diffraction pattern.*

Fig. 1 outlines the workflow for synthesizing a high-fidelity diffraction pattern from a raw multi-exposure stack. As illustrated in the optical schematic (Fig. 1a), the probe beam scans the sample plane to record diffraction intensities. To capture the full spatial frequency spectrum, data is acquired at $K$ distinct exposure times $t_k$ for each scanning position $\boldsymbol{r}_j$. Short exposures record the unsaturated low-frequency envelope, while long exposures resolve weak high-frequency scattering signals. The raw measurement at position $j$ is defined as the bracketed dataset $\mathcal{S}_j$:

$$\mathcal{S}_j = \{(I_{j,k}, t_k) \mid k = 1, \dots, K\},$$

where $I_{j,k}$ denotes the diffraction intensity at position $j$ with exposure time $t_k$.

The fusion process begins by evaluating the reliability of each frame $I_{j,k}$ in $\mathcal{S}_j$ (Fig. 1b). A non-linear mapping function assigns a pixel-wise confidence score $W_k$ based on local exposure quality (detailed in Methods, Eq. 5). This step acts as a spatial band-pass filter, retaining well-exposed regions while suppressing noise-dominated and saturated pixels. To prevent fusion artifacts, the signal is processed in the transform domain (Fig. 1c). Each diffraction pattern $I_{j,k}$ is decomposed into a Laplacian pyramid $\{L_l[I_{j,k}]\}$ to separate structural features across spatial scales (Methods, Eq. 7). Concurrently, the associated weight maps are smoothed into Gaussian pyramids $\{G_l[W_k]\}$ to match the feature resolution. In the final stage (Fig. 1d), the Laplacian features from $\mathcal{S}_j$ are modulated by their local confidence weights and quantized to filter incoherent noise. The fused pyramid is then collapsed to reconstruct the final pattern $I_{\mathrm{MNF}}$. This approach maintains the global physical consistency necessary for phase retrieval while improving the local SNR.

This framework provides a physics-aware regularization mechanism. By integrating the most reliable spectral components from each exposure, Multi-Scale Non-Linear Fusion (MNF) addresses the dual requirements of dynamic range expansion and noise suppression. The hierarchical decomposition preserves the global intensity envelope within the low-frequency base layers, which is necessary for satisfying the overlap constraint. Concurrently, the quantization-driven fusion filters high-frequency shot noise. Ultimately, this method converts photon-limited measurements into physically consistent, high-SNR inputs, facilitating stable convergence and improved resolution in the subsequent phase retrieval steps.

## 2.2 Performance of different methods in ptychography experiments

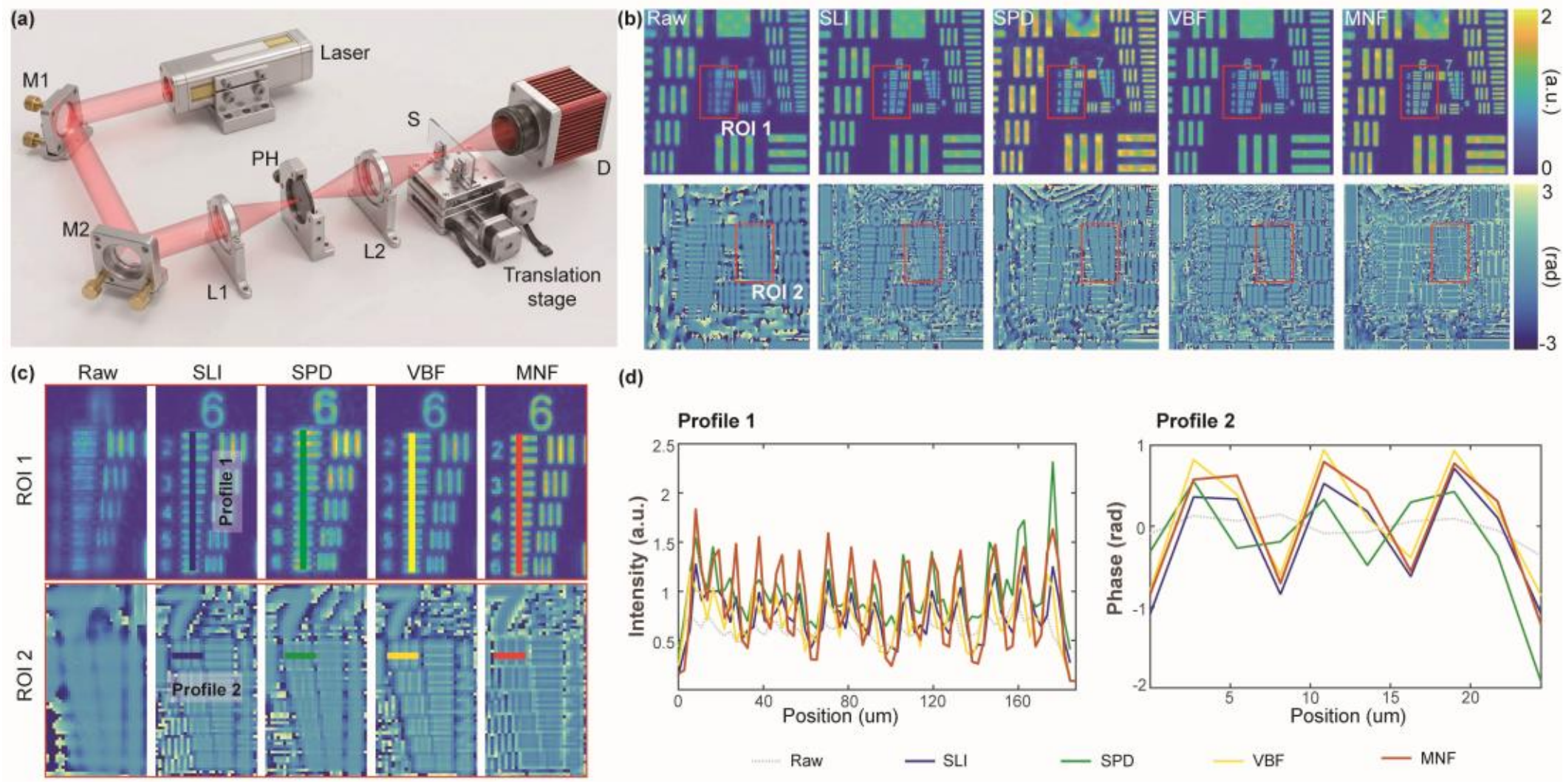

*Fig 2. Experimental validation of the proposed fusion framework using quasi-monochromatic illumination. (a) Schematic of the optical setup. A Helium-Neon laser serves as the illumination source, and the diffraction patterns are recorded by a detector in transmission geometry. (b) Comparison of reconstructed complex object functions (amplitude and phase) obtained via Single Exposure (Raw), Standard Linear Integration (SLI), Structural Patch Decomposition (SPD), Variance-Weighted Bayesian Fusion (VBF), and the proposed MNF. Red rectangles indicate the regions of interest (ROIs) selected for detailed inspection. (c) Magnified views of the high-frequency regions (ROI 1 for amplitude, ROI 2 for phase) near the Rayleigh resolution limit. The colored lines mark the paths used for cross-sectional analysis. (d) Quantitative line profiles extracted from the paths in (c). Profile 1 (Amplitude) and Profile 2 (Phase) demonstrate that MNF (orange curve) achieves the optimal*

*balance between noise suppression and contrast preservation, effectively resolving fine features that are obscured by noise in SLI or blurred by over-smoothing in SPD.*

We evaluated the proposed MNF strategy against established methods, including Standard Linear Integration (SLI), SPD, and Variance-Weighted Bayesian Fusion (VBF), using the experimental setup shown in Fig. 2a. The reconstructed complex object functions (Fig. 2b) show clear differences in amplitude and phase distributions, particularly in image contrast. While the SPD method produces the highest visual contrast, an examination of the high-frequency region near the Rayleigh resolution limit (Fig. 2c) highlights differences in reconstruction fidelity. The MNF method yields better intensity spectrum fitting and phase retrieval accuracy, resolving fine features that appear blurred or are lost with the other techniques.

Throughout the reconstruction process, we monitored quantitative metrics to benchmark these strategies (Fig. 2d). Consistent with the visual results, MNF achieves the most stable convergence and highest structural fidelity. Although SPD yields high numerical contrast in background regions, its reconstructed phase values deviate from expected physical quantities, indicating a loss of accuracy due to over-smoothing. VBF performs well generally but retains residual noise artifacts. SLI, limited by the direct propagation of shot noise, provides only marginal improvement over a single-exposure baseline.

A detailed analysis of optimization dynamics and signal characteristics reveals the physical mechanism underlying the MNF method's performance (Supplementary Notes 2–4). The error convergence trajectories show differences in how the strategies traverse the optimization landscape. Linear methods (SLI) and parametric models (VBF) exhibit a rapid initial descent (Supplementary Fig. 2), which we attribute to overfitting to noise. In contrast, MNF displays a more gradual convergence profile because the non-linear fusion weights act as a spatially adaptive spectral preconditioner. We also evaluated the trade-off between physical linearity and noise suppression. Quantitative analysis (Supplementary Fig. 3) shows that MNF operates in a regime with a moderate radiometric deviation of approximately 13% relative to the linear baseline. By doing so, MNF achieves a low noise floor while preserving the high-frequency diffraction peaks necessary for high resolution (Supplementary Fig. 4), effectively separating physical features from stochastic fluctuations. Furthermore, algorithmic stability analysis (Supplementary Note 4) indicates that reconstruction fidelity remains consistent across various hyperparameter settings, with relative error variations below 9%. Together with a linear computational complexity of $\mathcal{O}(N)$, this stability makes MNF a practical approach for high-throughput nanoscale imaging.

## 2.3 Broadband diffractive imaging with HHG XUV source

Synchrotron facilities provide exceptional brilliance for X-ray imaging, but their large infrastructure and limited beamtime restrict broader application. Integrating tabletop broadband XUV or X-ray sources with ptychography offers a more accessible approach to nanoscale imaging. We demonstrated the adaptability of our method to these laboratory-scale sources using a tabletop XUV setup (the emission spectrum and optical layout are shown in Fig. 3a; experimental parameters are detailed in Methods Section 3.2). Because the restricted spectral window (46–58 nm) of this source made it difficult to find samples with matched absorption edges, we measured standard semiconductor structures in a grazing incidence geometry (78°). This configuration confirms that our approach improves reconstruction fidelity under discrete broadband XUV illumination. Furthermore, the incident XUV beam in this experiment contains two distinct harmonics, which degrades temporal coherence. To our knowledge, this is the first experimental demonstration of HDR ptychographic reconstruction using such multi-harmonic, temporally incoherent diffraction data (Dierolf et al. 2010; Liu et al. 2021; Kodgirwar et al. 2024; Takahashi et al. 2023; Thibault and Guizar-Sicairos 2012; Wei et al. 2022; Ma et al. 2017; Xu et al. 2022).

The reconstructed phase maps (Fig. 3b) show that the MNF method avoids the over-smoothing seen in conventional approaches. It effectively recovers high-frequency spatial features, sharp edges, and distinct phase boundaries. We analyzed the 1D phase profiles and spatial resolution metrics (Supplementary Fig. 6) to quantify these differences. Cross-sectional phase profiles, extracted along the solid line in Supplementary Fig. 6a, are presented in Supplementary Fig. 6b. While the SLI, VBF, and SPD methods show substantial over-smoothing and phase underestimation, MNF accurately preserves steep edge transitions. We further evaluated spatial resolution using the Modulation Transfer Function (MTF) (Supplementary Fig. 6c). The MNF method maintains MTF values above 0.5 (the 50% threshold) across the frequency spectrum up to the Nyquist limit (0.5 cyc/pixel).

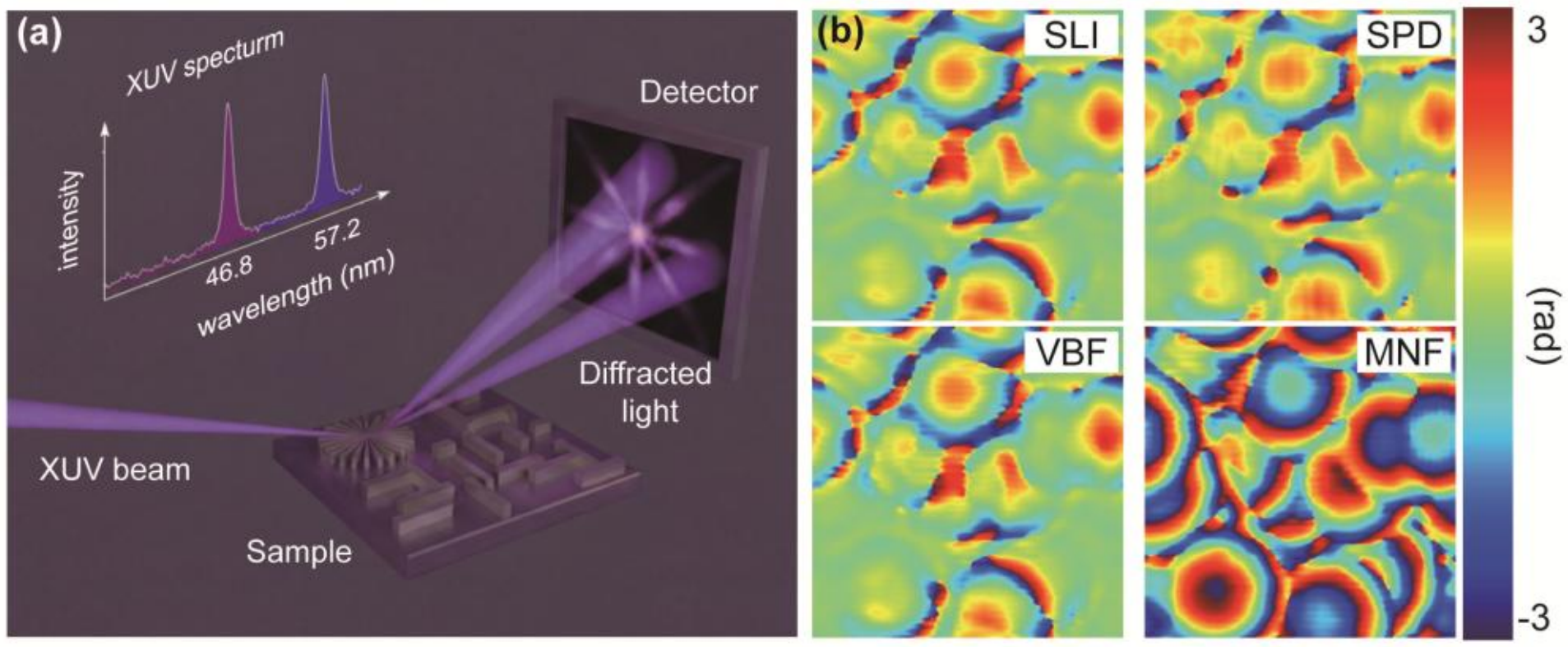

*Fig 3. Phase reconstruction under multi-harmonic tabletop XUV illumination. (a) Experimental layout in a grazing incidence geometry, with the inset showing the*

*discrete broadband XUV spectrum. (b) Reconstructed phase maps. The MNF method successfully recovers sharp structural edges, outperforming conventional approaches (SLI, SPD, VBF) which suffer from severe over-smoothing.*

To illustrate the challenges of acquiring high-quality diffraction data under broadband XUV illumination, Figs. 4a–c show the impact of varying exposure times (1 s, 2 s, and 4 s) on the raw diffraction patterns. Under a short 1-s exposure (photon-starved regime, Fig. 4a), the diffraction signal is extremely weak. Extending the exposure to 4 s significantly enhances the central diffraction spot but inevitably induces severe detector saturation. Figure 4b quantifies the noise variance within the structure-free background regions (yellow dashed boxes in Fig. 4a). Increasing the exposure from 1 s to 4 s amplifies the background noise variance by a factor of 80, indicating that prolonged exposure introduces detrimental background noise. Furthermore, the 1D intensity profiles extracted along the white solid lines (Fig. 4c) confirm the detector saturation effect. Unlike the 1-s and 2-s profiles, the 4-s exposure curve exhibits a pronounced flat-top truncation at the peak. This not only distorts the low-frequency signals but also elevates the overall baseline, thereby submerging the weak high-frequency diffraction information. This quantitative analysis demonstrates that, constrained by the limited dynamic range of the detector, a single exposure cannot simultaneously capture both the intense central beam and the faint peripheral diffraction signals. This fundamental limitation provides the physical premise for implementing HDR fusion techniques.

Accordingly, we evaluated four different HDR fusion algorithms to overcome this bottleneck. As visually compared in the diffraction patterns in Fig. 4d, the conventional SLI and SPD methods still suffer from obvious residual background noise. While the VBF method effectively suppresses this noise, it severely attenuates the intensity of the central valid signal. In contrast, our proposed MNF method maintains excellent background noise suppression while fully preserving the intense central signal. This visual fidelity is corroborated by the SNR analysis of a single diffraction pattern, as shown in Fig. 4e. For this selected frame, the MNF method achieves the highest SNR of 31.66 dB, significantly outperforming the other conventional approaches.

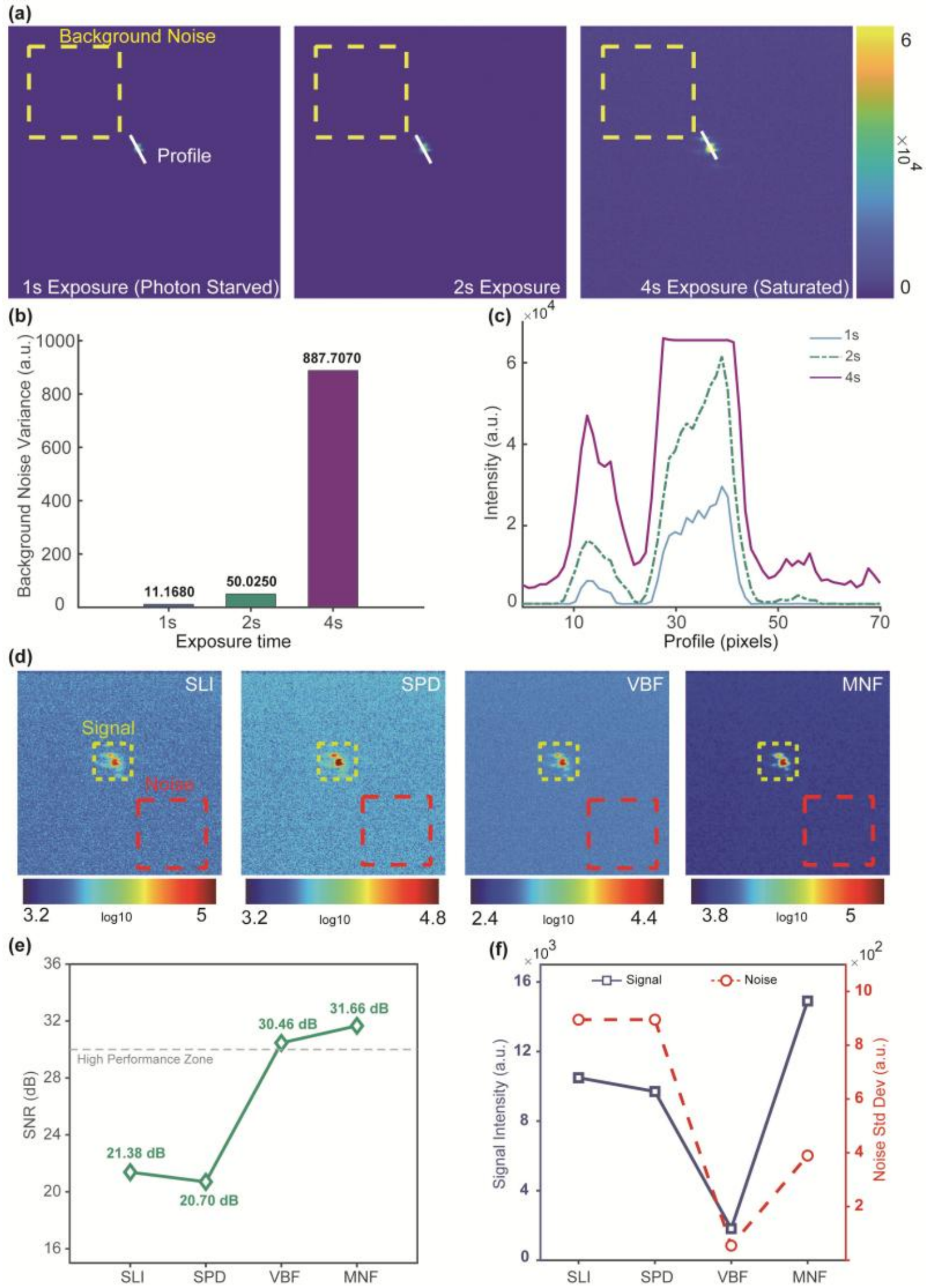

*Fig 4. Experimental validation of the HDR fusion framework under broadband XUV illumination. (a) Raw diffraction patterns acquired at 1 s, 2 s, and 4 s exposure times. (b) Background noise variance extracted from the structure-free regions (yellow dashed boxes in (a). (c) 1D intensity profiles plotted along the white solid lines in (a). (d) HDR diffraction patterns fused by the SLI, SPD, VBF, and proposed MNF methods. (e) Signal-to-noise ratio (SNR) evaluation for a representative single diffraction frame. (f) Extracted signal intensity and noise standard deviation for the four fusion algorithms.*

# 3 Methods

## 3.1 Details of the optical experiment

Quasi-monochromatic illumination was provided by a Helium-Neon laser (HNL050LB, Thorlabs). The beam was collimated by mirrors M1 and M2 and subsequently focused onto a pinhole aperture (PH) via a lens (L1) with a focal length of 50 mm. An imaging lens (L2, $f = 40$ mm) relayed the pinhole to form a probe spot approximately 100 $\mu$m in diameter on the sample plane. This focused probe interacted with the specimen in transmission geometry, generating coherent diffraction patterns. The sample (S) was translated along a spiral scanning trajectory using linear stages. To ensure sufficient data redundancy for robust phase retrieval, a step size of 20 $\mu$m was employed, yielding an overlap of approximately 80% between adjacent scan positions. The resulting diffraction intensities were recorded by a scientific complementary metal oxide semiconductor (CMOS) detector (QHY600PH, QHYCCD) positioned 16.5 mm downstream from the sample.

## 3.2 Details of the XUV Experiment

The tabletop HHG source was driven by a commercial femtosecond laser (Pharos, Light Conversion) operating at a central wavelength of 1030 nm. The fundamental beam was frequency-doubled through a 0.5-mm BBO crystal to generate 515-nm pulses with an average power of 2 W and a pulse duration of 250 fs. This driving laser was focused into an argon-filled gas cell, consisting of a capillary with an inner diameter of 150 $\mu$m and an outer diameter of 1200 $\mu$m, for HHG. To isolate the XUV radiation, two 200-nm-thick aluminum filters were employed to block the residual driving laser. The resulting HHG beam, containing two dominant harmonic components, was focused by a toroidal mirror onto a 50-$\mu$m pinhole to define the probe. The sample was mounted on a vacuum translation stage, and the exit wave propagated in free space over a distance of 50 mm before being recorded by a detector (XF95, Tucsen) with a pixel pitch of 11 $\mu$m. This setup yielded 512×512-pixel polychromatic diffraction patterns, which were directly processed by the fusion algorithms without prior numerical monochromatization.

## 3.3 Physical forward model

In far-field ptychography, the interaction between the probe beam and the specimen is modeled under the projection approximation. The complex exit wave $\psi_j(\boldsymbol{r})$ at a scanning position $\boldsymbol{r}_j$ is defined as the product of the probe function $P(\boldsymbol{r})$ and the object transmission function $O(\boldsymbol{r})$:

$$\psi_j(\boldsymbol{r}) = P(\boldsymbol{r})O(\boldsymbol{r} - \boldsymbol{r}_j),$$

where $\boldsymbol{r}$ denotes the lateral position vector in the sample plane. Under the Fraunhofer approximation, the ideal diffraction intensity $I_j^{\text{coh}}(\boldsymbol{q})$ recorded in the far-field is proportional to the squared modulus of the Fourier transform of the exit wave:

$$I_j^{\text{coh}}(\boldsymbol{q}) = \left|\mathcal{F}[\psi_j(\boldsymbol{r})]\right|^2,$$

where $\boldsymbol{q}$ represents the spatial frequency coordinate in the detector plane, and $\mathcal{F}$ denotes the Fourier transform operator.

In experimental scenarios, the measured intensity $I_j^{\text{meas}}(\boldsymbol{q})$ accounts for both the spectral properties of the source and stochastic detection noise $\eta(\boldsymbol{q})$. We model this as a generalized spectral superposition:

$$I_j^{\text{meas}}(\boldsymbol{q}) = \int S(\lambda)\left|\mathcal{F}[P_\lambda(\boldsymbol{r})O_\lambda(\boldsymbol{r}-\boldsymbol{r}_j)]\right|^2 d\lambda + \eta(\boldsymbol{q}),$$

where $S(\lambda)$ represents the spectral power density. This formulation encompasses two distinct regimes:

- **Quasi-monochromatic regime:** For narrow-band sources (e.g., lasers), $S(\lambda) \approx \delta(\lambda - \lambda_0)$. The measurement approximates the ideal coherent intensity (Eq. [eq:coherent_intensity]), where the primary challenge is the limited dynamic range of the detector.
- **Broadband regime:** For polychromatic sources (e.g., HHG), the integration over $S(\lambda)$ leads to radial smearing of high-frequency fringes. Here, the challenge is compound: dynamic range limitations coupled with spectral decoherence.

## 3.4 Detailed implementation of multi-Scale non-linear fusion

The MNF strategy processes the raw measurement $I_j^{\text{meas}}$ to synthesize a high-fidelity diffraction pattern $I_{\text{MNF}}$. For each scanning position $j$, the measurement consists of a bracket of $K$ exposures $\{I_{j,k}\}_{k=1}^{K}$ acquired with varying exposure times. The algorithm fuses this set through four sequential stages performed in the reciprocal space coordinate $\boldsymbol{q}$.

### *3.4.1 Exposure quality assessment.*

For each exposure $k$, a pixel-wise weight map $W_k$ quantifies information reliability. The normalized intensity $\bar{I}_{j,k}(\boldsymbol{q}) \in [0,1]$ serves as the input. The weight is modeled as a Gaussian likelihood function:

$$W_k(\boldsymbol{q}) = \exp\left(-\frac{(\bar{I}_{j,k}(\boldsymbol{q}) - \mu)^2}{2\sigma^2}\right),$$

where $\mu$ is the optimal dynamic range midpoint, and $\sigma$ controls the selectivity of the weighting curve. This function assigns high weights ($W_k \to 1$) to pixels within the linear response region, while penalizing under-exposed and saturated regions. The weights are subsequently normalized such that $\sum_k W_k(\boldsymbol{q}) = 1$.

### 3.4.2 Multiresolution pyramid decomposition.

To enable scale-dependent processing, both the image $I_{j,k}$ and its weight map $W_k$ are decomposed into multiresolution pyramids with $L$ levels. Two Gaussian pyramids are generated: one for the image intensity $\{G_l[I_{j,k}]\}$ and one for the weight map $\{G_l[W_k]\}$. The decomposition follows the recursive filtering rule:

$$G_{l+1}[\cdot] = \mathcal{D}_{\downarrow 2}(G_l[\cdot] * w_G),$$

where $*$ denotes convolution with a Gaussian smoothing kernel $w_G$, and $\mathcal{D}_{\downarrow 2}$ represents the spatial down-sampling operator. The structural information of the image is further extracted into a Laplacian pyramid $L_l[I_{j,k}]$. For each level $l < L$, the band-pass features are obtained by subtracting the up-sampled coarser layer from the current Gaussian layer:

$$L_l[I_{j,k}](\boldsymbol{q}) = G_l[I_{j,k}](\boldsymbol{q}) - \mathcal{U}_{\uparrow 2}\big(G_{l+1}[I_{j,k}]\big)(\boldsymbol{q}),$$

where $\mathcal{U}_{\uparrow 2}$ denotes the up-sampling operator followed by interpolation filtering. This step isolates fine diffraction fringes (high $l$) from the coarse intensity envelope (low $l$).

### 3.4.3 Coefficient modulation and quantization.

Fusion is performed in the transform domain for each exposure individually. The Laplacian coefficients are first modulated by their corresponding weight map: $\hat{L}_l^{(k)} = G_l[W_k] \cdot L_l[I_{j,k}]$. To suppress noise, we apply a scale-adaptive quantization constraint $\mathcal{Q}(\cdot)$ to these modulated coefficients:

$$\tilde{L}_l^{(k)}(\boldsymbol{q}) = \Delta_l \cdot \text{round}\left(\frac{\hat{L}_l^{(k)}(\boldsymbol{q})}{\Delta_l}\right).$$

The quantization step $\Delta_l$ is adaptive to the pyramid level, typically decreasing from coarse to fine scales to account for the spectral decay of the signal. This operation acts as a discrete denoising filter applied individually to each exposure channel, enforcing sparsity on the high-frequency coefficients.

### 3.4.4 Reconstruction and accumulation.

Instead of fusing coefficients into a single pyramid, we reconstruct a "filtered" image $R_k$ for each exposure $k$. The reconstruction proceeds recursively from the coarsest level $L$ to the finest level 0. Importantly, the base structural information is propagated from the Gaussian pyramid, while the weight modulation is selectively applied to the detail layers:

$$R_{l-1}^{(k)}(\boldsymbol{q}) = \mathcal{U}_{\uparrow 2}(G_l^{(k)})(\boldsymbol{q}) + \tilde{L}_{l-1}^{(k)}(\boldsymbol{q}),$$

where $\mathcal{U}_{\uparrow 2}(G_l^{(k)})$ represents the up-sampled Gaussian image from the coarser level (acting as the structural base), and $\tilde{L}_{l-1}^{(k)}$ is the quantized, modulated detail coefficient. Finally, the total fused diffraction pattern $I_{\mathrm{MNF}}$ is obtained by accumulating the reconstructed contributions from all exposures:

$$I_{\mathrm{MNF}}(\boldsymbol{q}) = \sum_{k=1}^{K} R_0^{(k)}(\boldsymbol{q}).$$

This hybrid strategy ensures that the low-frequency intensity envelope (carried by the Gaussian base) is preserved to maintain physical consistency, while high-frequency noise is rigorously filtered via the weighted quantization of the Laplacian details.

The datasets generated and analyzed during the current study are not publicly available due to project restrictions but are available from the corresponding author upon reasonable request.

Codes used to post-process the diffraction data within this paper are available from the corresponding author upon request.

The authors declare no competing financial or non-financial interests.

References

Y.X. and S.F.S. conceived the research and designed the non-linear spectral preconditioning framework. Y.X. developed the algorithms, performed numerical simulations, and processed experimental data. Y.X., B.Z., and Y.Z. conducted the XUV imaging experiments. Z.H. provided technical support for ptychographic reconstruction. Y.X. wrote the manuscript with contributions from all authors. S.F.S. oversaw the project.